\RequirePackage{fix-cm}
\documentclass[twocolumn]{svjour3}          
\smartqed  
\usepackage{graphicx}
\usepackage[utf8]{inputenc}
\usepackage{xcolor}
\usepackage{natbib}
\usepackage{lineno}
\usepackage{colortbl}
\begin{document}

\title{Is the faint young Sun problem for Earth solved?
}


\author{Benjamin Charnay \and
  Eric T. Wolf \and
  Bernard Marty \and 
  François Forget}


\institute{B. Charnay \at
              LESIA, Observatoire de Paris, Université PSL, CNRS, Sorbonne Université, Université de Paris, 5 place Jules Janssen, 92195 Meudon, France. \\
              \email{benjamin.charnay@obspm.fr}           
           \and
           E. T. Wolf \at
           University of Colorado, Boulder, Laboratory for Atmospheric and Space Physics, Department of Atmospheric and Oceanic Sciences, Boulder, CO 80302, USA.
           \and   
           B. Marty \at
              Centre de Recherches Pétrographiques et Géochimiques, UMR 7358 CNRS – Université de Lorraine, 15 rue Notre Dame des Pauvres, BP 20, 54501 Vandoeuvre-lès-Nancy, France.
           \and
           F. Forget \at
             Laboratoire de M\'et\'eorologie Dynamique/IPSL, CNRS, Sorbonne Universit\'e, Ecole Normale Sup\'erieure, PSL Research University, Ecole Polytechnique, 75005 Paris, France.
              }

\date{Received: date / Accepted: date}

\maketitle

\begin{abstract}

Stellar evolution models predict that the solar luminosity was lower in the past, typically 20-25 $\%$ lower during the Archean (3.8-2.5 Ga). Despite the fainter Sun, there is strong evidence for the presence of liquid water on Earth's surface at that time. This “faint young Sun problem” is a fundamental question in paleoclimatology, with important implications for the habitability of the early Earth, early Mars and exoplanets. 
Many solutions have been proposed based on the effects of greenhouse gases, atmospheric pressure, clouds, land distribution and Earth’s rotation rate.
Here we review the faint young Sun problem for Earth, highlighting the latest geological and geochemical constraints on the early Earth's atmosphere, and recent results from 3D global climate models and carbon cycle models. 
Based on these works, we argue that the faint young Sun problem for Earth has essentially been solved. Unfrozen Archean oceans were likely maintained by higher concentrations of CO$_2$, consistent with the latest geological proxies, potentially helped by additional warming processes. This reinforces the expected key role of the carbon cycle for maintaining the habitability of terrestrial planets.
Additional constraints on the Archean atmosphere and 3D fully coupled atmosphere-ocean models are required to validate this conclusion.

\end{abstract}

\section{Introduction}
\label{intro}

Let's imagine time-travelers exploring the early Earth 3-4 billions years ago. Leaving their time capsule, they would need oxygen masks to survive in the anoxic Archean atmosphere, containing deathly high levels of carbon dioxide and possibly also methane, carbon monoxide, ammonia and hydrocyanic acid. Moreover, with no appropriate protection, they would quickly suffer from sunburn, due to the higher UV flux of the early Sun and the absence of an ozone layer. Passed these complications, they would then discover a world very different from our present-day Earth. The sky would look hazier, potentially with an orange colour. The days would be shorter and a larger moon would shine in the nightsky. The Sun would appear significantly fainter than today. However, perhaps surprisingly, most of the surface would be covered by temperate or warm liquid water oceans. These oceans would be more acidic and full of unicellular organisms forming a productive biosphere. Our time-travelers may wonder how can the climatic conditions be clement under such a fainter Sun. Theywould experience the faint young Sun problem, one of the most fundamental questions in paleoclimatology.

The faint young Sun problem for Earth has been reviewed in detail by \cite{feulner12}. At that time, most of the atmospheric modelling work was based on 1D radiative convective models (RCMs). 
While 1D RCMs remain attractive research tools due to their flexibility and computational efficiency, 1D models miss important climate feedbacks.  For instance, RCMs used in early Earth and habitability studies typically omit explicit representations of clouds and surface ice and snow.  Instead, the surface albedo is assigned a constant high value, to implicitly represent cloud reflectivity.  By treating the radiative effect of clouds in this manner, such models fail to include the longwave radiative effect of clouds and can overestimate the greenhouse effect of background gases \citep{pierrehumbert95,goldblatt11}. They also miss cloud feedbacks related to changes in the cloud distribution.

Three-dimensional atmospheric general circulation models (GCMs) coupled to simple ocean/sea-ice models represent a significant improvement in the simulation of the climate system compared to 1D RCMs.  GCMs allow for a self-consistent and coupled treatment of numerous dynamical and physical processes occurring in planetary atmospheres.  Of particular importance for Earth’s climate system, is the treatment of water in its various thermodynamic phases.  
In particular, water is responsible for the sea-ice albedo, water vapor greenhouse and cloud feedbacks, which can all strongly impact the climate sensitivity.
Shortly after the review from \cite{feulner12}, several papers using 3D atmospheric GCMs coupled to simple ocean/sea-ice models were published, providing a fresh new perspective and some answers about former proposed solutions to the faint young Sun problem \citep{charnay13, wolf13, wolf14,  lehir14, kunze14, teitler14, charnay17, wolf18}. According to these modelling studies, the faint young Sun problem appears less severe than initially thought, notably because of cloud feedbacks leading to optically thinner low clouds on the early Earth.

In parallel, new geological and geochemical constraints were obtained during the last decade, changing the general view about the evolution of Earth's atmospheric nitrogen, as well as constraints on atmospheric CO$_2$ and surface temperatures during the Archean. In particular, some of these recent studies suggest relatively high pCO$_2$ values, which mitigate the faint young Sun problem.

This paper constitutes a general review of the faint young Sun problem for Earth, highlighting atmospheric modelling results and some laboratory measurements obtained during the last decade.
In section 2, we describe the environment of the early Earth, in particular the evidence for a weaker Sun, for the presence of surface liquid water, and for the fraction of emerged land. We also detail constraints on the ocean temperature and atmospheric composition/pressure.
Section 3 lists different proposed solutions to the faint young Sun problem, including the effects of greenhouse gases, atmospheric pressure, clouds and the fraction of emerged land. For each solution, we highlight recent results from climate modelling and paleosol studies.
Section 4 reviews the lessons from 3D global climate models and the differences compared to 1D atmospheric models.
Section 5 discusses the role of the carbon-cycle and life to regulate the early Earth's climate.
Finally, we conclude in section 6 with a summary and possible directions for future research.

\section{The environment of the early Earth}
\subsection{Evidence for a weaker Sun}

It has been well established in models of solar evolution that solar luminosity has slowly increased over the lifetime of the Sun. 3.8 billions years ago the early solar flux is calculated to have been 25$\%$ lower than today and it has slowly intensified since (see e.g. \cite{newman77, gough81}). This is because the fusion of hydrogen into helium increases the mean molecular weight of the core. To maintain the balance between the pressure gradient force and gravity, the core contracts and warms. The increased densities and temperatures enhance the rate of fusion and hence, the star's luminosity increases with time.  This model is very robust.
Furthermore, it is based on a model of the Sun which agrees very well with solar neutrinos measurements and helioseismology observations. 

Can this model be questioned? Is it possible to imagine a young Sun brighter than expected that could resolve the faint young Sun problem and possibly explain the very warm temperature that have been reported by some studies? More than for the Earth, such a possibility has been discussed to resolve the enigma of early Mars covered by rivers and lakes 3.5 - 4 billions years ago \citep{whitmire95, forget13, haberle17}.

The only way to explain a brighter early Sun while remaining consistent with the robust theory mentioned above is to assume that it was more massive initially and that it subsequently lost the excess mass in a solar wind much more intense than today \citep{whitmire95}. Solar luminosity is proportional to the fourth power of solar mass $M_{\odot}$, and since a planet's orbital distance $r$ is inversely proportional to $M_{\odot}$, and the solar flux varies as $r^{-2}$, the flux at a planet scales as $F \sim M_{\odot}^6$ \citep{whitmire95}.
Therefore a few percent of the mass would be sufficient. An initial mass of $\sim 6$\% higher than today makes the Sun as bright as today 4.5 Ga.

For years, this assumption has been regarded with skepticism by the community, because of the lack of evidence and the success of the "Standard Solar Model" of the Sun. However, a decade ago some inconsistencies between the standard solar model predictions and measurements of the solar CNO abundances were revealed \cite{asplund09, caffau08, caffau10}. Interestingly a younger more massive Sun was then suggested to explain these inconsistencies \citep{guzik10, turck-chieze11}. 
Overall assuming an early massive Sun has consequences on the modelled solar evolution and leaves traces on the numerous observational constraints that are available for the Sun itself (seismic properties, neutrino fluxes, composition). Yet the most recent studies on the subject conclude that these datasets cannot rule out the early massive Sun hypothesis (See \cite{wood18, buldgen19}, and references therein). 

The intensity and the evolution of the early solar wind remain open questions. Unfortunately, the basic mechanisms responsible for producing winds from solar-like main sequence stars are still not understood well enough to provide quantitative information. Ideally, much could be learned by observing stellar winds around other stars at various ages. However stellar winds from solar-mass stars are very difficult to observe directly owing to their low optical depth. Some studies have been performed by looking for the Ly$\alpha$ signature of the charge exchange that occurs when the ionized stellar wind collides with the neutral interstellar medium \citep{zank99, wood02, wood05}. Analysing these observations for stars of various ages suggest that stellar winds may decrease exponentially with time and that most of the extra mass was lost in less than a few hundreds of million years \citep{wood05, minton07}. 
Observations of the radio emission of young solar-type stars suggest a total solar mass loss lower than 2$\%$ after 100 Myr \citep{fichtinger17}. Finally, observations of stellar spin down rate have been used to argue in favour of a large and sustained mass loss and a more massive young Sun \citep{martens17}, but they also suggest an exponential decrease of the mass loss \citep{gallet13}. 

Therefore the key difficulty in solving the faint young Sun problem with a more massive sun is to keep the Sun sufficiently massive and bright for one or two billion of years, throughout the Archean. Observations of young solar analogs tend to rule out this possibility, although more data are required to be certain \citep{wood18}. In the future, it may be possible to constrain the Sun's ancient mass and its evolution by looking for remaining signatures of the early solar wind and the effects of a solar mass-loss history on the orbital dynamics in the planetary system, including possible signatures in the geological and climate record \citep{minton07, spalding18}.

\subsection{Evidence for liquid water}
The oceans might have formed rapidly in a gigantic deluge, a few Myr after the cataclysm that led to the Earth-Moon system, 4.5 Gyr ago \citep{sleep01b}. Some models predict that oceans could have formed even earlier and could have partly survived the Moon forming event \citep{genda05}. The first evidence for the presence of liquid water on Earth derives from the analysis of fragments of Archean zircons. Zircons are U-rich magmatic minerals that are resilient enough to have survived several mountain building cycles. Some of them contain fragments having U-Pb ages up to 4.2-4.4 Ga \citep{wilde01,mojzsis01}. These old zircons present oxygen isotopic compositions that cannot be explained by a dry magmatic genesis and require contribution of material having been hydrothermally altered at low temperature. Direct geologic evidence of oceans stems from the occurrence of pillow basalts and layered banded iron formations in 3.8 Ga units, demonstrating underwater magmatism and sedimentation at, or prior to, that time. Well preserved formations with low metamorphic grades abound 3.5 Ga ago in NW Australia and South Africa. Hydrothermal quartz at that ages contain fluid inclusions which contain mixtures of hydrothermal end-member(s) and Archean seawater \citep{foriel04}. Coupled Cl-K-Ar study suggests a salinity comparable to the modern one and, possibly temperatures in the range 20-40$^\circ$C (see below) \citep{marty18}. Both thermal modelling and field evidence are consistent with the occurrence of liquid water on Earth within a few tens of Ma after the formation of the solar system, with direct evidence for oceans at 3.8-3.5 Ga.

\subsection{Constraints on the fraction of land}
The fraction of emerged land affects the surface albedo and thus the planetary energy budget.
The emergence of land required the development of continents able to float over the denser mantle. Hence the continental lithosphere needed to be strong enough to support crustal thickening and high reliefs \citep{rey08}. The period of time when this happened is difficult to estimate and models are generally based on the distribution of old continental crust terrains, as well as on the petrology and geochemistry of these geological
units \citep{allegre84,belousova10,dhuime12,pujol13}. These models propose major pulses of crustal generation in the period 4 Ga - 3 Ga that resulted in 60-80$\%$ present-day volume at 3 Ga (see  Figs. 5 \& 6 in \cite{hawkesworth19}). After 3 Ga, the net growth rate diminished because crustal destruction became effective and competed efficiently with crustal generation. This period of time may mark the onset of modern-style plate tectonics. A lower continental crust volume combined with potentially a larger ocean volume \citep{pope12} logically implies less emerged land during the Archean than today. \cite{flament08} estimated that the fraction of emerged land was lower than 12$\%$ and probably around 2-3$\%$ (compared to 27$\%$ today).

\subsection{Geological constraints on the temperature}
A long debate is still ongoing about the temperature of the early oceans. Archean oceanic cherts (SiO$_2$) appear depleted in $^{18}$O relative to $^{16}$O \citep{knauth03, robert06, tartese16}. A possible explanation of this trend is related to the temperature of Archean oceans. Indeed, cherts concentrate less $^{18}$O relative to seawater during their formation as oceanic temperature increases. The low $\delta^{18}$O have thus been interpreted to suggest hot oceans with temperatures between 60$^\circ$C (333 K) and 80$^\circ$C (353 K) \citep{knauth03, robert06, tartese16}. Notably, silicon isotopes coupled with oxygen isotopes also attest for oceanic temperatures warmer than today \citep{robert06}. The temperature range inferred from Archean cherts is compatible with the thermophiles inferred from evolutionary models of the ancient life \citep{gaucher08}, with constraints on the salinity of Archean oceans \citep{marty18} and possibly with the indications for a low ocean water viscosity \citep{fralick11}. If the interpretation of low $\delta^{18}$O in Archean cherts is correct, such high oceanic temperatures would make the faint young Sun problem for Earth much more difficult to overcome (see section 4.2 and 5).

However, the interpretation of $\delta^{18}$O isotope ratio as indicator of warm Archean oceans has been strongly debated. Some analyses suggest temperate oceans with temperatures lower than 40$^\circ$C \citep{hren09, blake10}. In addition, most of Archean cherts did not directly precipitate from seawater or may not be sufficiently well preserved to be used to reconstruct past oceanic temperature \citep{marin-carbonne12, marin-carbonne14, cammack18}. Other possible explanations have been proposed as hydrothermal alteration of the seafloor \citep{vandenboorn07} or as a gradual changes in the oxygen isotope composition of seawater \citep{kasting06a, jaffres07}. That latter criticism has been ruled out by \cite{tartese16}, who investigated O isotopes not in cherts but in kerogen and demonstrated that the ocean $\delta^{18}$O composition remained almost constant. The triple oxygen isotope mass balance model from \cite{sengupta18} suggests that Archean cherts precipitated in cool oceans with modern-like $\delta^{18}$O followed by diagenetic alteration.

In addition, warm/hot Archean oceans are difficult to reconcile with the glacial evidence. The glacial Archean record includes the Huronian glaciations at 2.4 Ga and glacial rocks at 3.5, 2.9 Ga and 2.7 Ga \citep{kasting06a, ojakangas14, dewit16}. The Huronian glaciations were likely Snowball Earth events, where the entire planet was encases in ice,  while the others were likely only partial glaciations. They imply global mean temperatures below 20$^\circ$C \citep{dewit16}, at least episodically. The transitions from warm climates with mean surface temperature around 60$^\circ$C to cold climates or Snowball-Earth events would imply major changes in the atmospheric composition and in the carbon cycle, implying unknown mechanisms. Such swings in the carbon cycle seem unlikely, although the glacial events may represent only a small fraction of the Archean period.

Finally, evolutionary models suggest thermophilic ancient life but also a mesophilic last universal common ancestor (LUCA) \citep{boussau08}. These molecular thermometers do not give information about the global mean surface temperature. Thermophily near the roots of the tree of life may reflect local warm environments (like hydrothermal vents) or survival after hot climates produced by large impacts \citep{boussau08, abramov09}.

We conclude that the debate about the temperature of the early oceans is not over, although the arguments for cold or temperate climates seem stronger now. We discuss later the implications of both cases in the context of the faint young Sun problem.

\subsection{Geological constraints on the atmospheric composition and pressure}
The atmospheric composition and pressure in the distant past are primarily estimated from models of the origin(s) of atmospheric volatiles and from models of atmospheric evolution upon interactions with outer space and exchanges with the solid Earth (see the review by \cite{catling20} and references therein). Noble gas isotope systematics indicate that the atmosphere is an ancient reservoir that was formed a few tens of Ma after start of solar system formation 4.56 Ga ago, by degassing of material that accreted to form our planet. The geological record of ancient sediments and their weathering profile is consistent with relatively low partial pressures of CO$_2$ (pCO$_2$ $<$ 30 mbar) for the late Archean \citep{sheldon06,driese11}. A recent technique involving the modelling of the aqueous chemistry in paleosols gives higher estimations for pCO$_2$ during the Archean (pCO$_2$=24–140 mbar at 2.77 Ga, 22–700 mbar  at 2.75 Ga and 45–140 mbar at 2.46 Ga) and a gradual decrease with time \citep{kanzaki15}. \cite{rosing10} derived a very low upper limit of $\sim$ 0.9 mbar from the coexistence of siderite and magnetite in Archean banded iron formation. That limit based on thermodynamic arguments has yet been questioned \citep{reinhard11}. Finally, \cite{lehmer20} interpreted the oxidation state of micrometeorites at 2.7 Ga as a possible evidence for a CO$_2$ atmospheric mixing ratio $>$ 70\%.
These geological constraints on pCO$_2$ are generally still debated, do not cover all times of the Archean, and are often incompatible. Therefore, we do not think that they should be considered as strict limits.
The geological record also suggests the onset of an oxygenated atmosphere around 2.4 Ga, called the Great Oxidation Event \citep{lyons14}. Removing CO$_2$ and O$_2$ from the atmospheric composition leaves N$_2$, water vapour, noble gases and several C species as the main atmospheric constituents. 

Geological measurements of the atmospheric pressure and compositions in the distant past were thought to be impossible due to the mobility of volatile elements in rocks and minerals undergoing metamorphism. Surprisingly, this assumption was not correct and new types and samples and new approaches permit measurable constraints to be determined. \cite{marty13} analyzed volatiles trapped in fluid inclusions in 3.5 Ga hydrothermal quartz from the Pilbara, NW Australia. $^{36}$Ar is a primordial isotope that has been conserved in the atmosphere, as indicated by constant $^{38}$Ar/$^{36}$Ar ratio. $^{40}$Ar is a radiogenic isotope produced by the decay of crustal and mantle $^{40}$K. The triple bond of N$_2$ makes this molecule very stable and N$_2$ is sometimes referred as “the sixth noble gas”. In a $^{40}$Ar/$^{36}$Ar versus N$_2$/$^{36}$Ar frame, these authors found that data from vacuum crushing experiments define a straight line representing mixing between a low $^{40}$Ar/$^{36}$Ar, N$_2$/$^{36}$Ar end-member representing paleo-atmospheric noble gases dissolved in seawater and an hydrothermal component rich in crustal $^{40}$Ar and N. The extrapolated seawater N$_2$/$^{36}$Ar ratio was comparable to, or lower than, the modern value, leading \cite{marty13} to propose that the pN$_2$ was $\leq$ 1.1 bar and possibly as low as 0.5 bar. The authors also found that the N isotopic composition of Archean air was similar to the modern composition,  thus discarding the possibility of isotopically fractionating atmospheric escape of nitrogen since 3 Ga. \cite{avice18} developed a similar study for several samples from another area (Berberton, South Africa) and proposed that the Archean pN$_2$ was  $\leq$ 0.5 bar. 
Independently, \cite{som12} attempted to set constraints on the barometric pressure in the distant past from fossil imprints of raindrops. The rationale is that the maximum size of raindrops is a function of the P$_{atm}$ as raindrop will fragment over a threshold size due to air drag. These authors measured 2.7 Ga imprints of raindrops and conducted in parallel analogic experiments to conclude that the Archean atmospheric pressure could not have been higher than 2.3 bar and was probably below 1.3 bar. This upper limit has been criticised by \cite{kavanagh15}, because the analysis of fossil imprints can be biased by very rare large raindrops. \cite{goosmann18} argued that such a bias is statistically unlikely given that 18 distinct bedding surfaces were analysed by \cite{som12}.

In a further study, \cite{som16} proposed an absolute barometric pressure of 0.23 $\pm$ 0.23 bar using the size distribution of gas bubbles in 2.5 Ga basaltic lava flows that solidified at sea level. The maximum size that bubbles can reach in a lava flow is a function of the external pressure on the surface of the flow, here the barometric pressure. They compared the bubble sizes for two different layers in the flow, at a given depth (which pressure can be evaluated from the weight of the lava column above) and at the flow’s surface, determined from morphological arguments. \cite{som16} suggested 0.5 bar as an upper limit for P$_{atm}$. Together, these studies suggest a surprisingly low barometric pressure for the Archean atmosphere, with little room for pCO$_2$, which given errors could not have been above 0.5 bar, and possibly in the range 0-0.2 bar. However, little is known about this long period of time for which sample ages cover about 0.8 Ga, and more detailed studies are necessary to better document the fate of atmospheric gases in the neo-Archean before the Great Oxidation Event.

\section{Proposed solutions to the faint young Sun problem}

With the solar constant $\sim$25$\%$ weaker at 3.8 Ga \citep{gough81}, the global mean insolation absorbed by the Earth had a deficit of $\sim$60 W/m$^2$ ($\sim$44 W/m$^2$ at 2.5 Ga) compared to the current value of 240 W/m$^{-2}$ \citep{wild13}. The Earth with the present atmospheric composition and continents would have undergone a cooling of $\sim$60 K (assuming the current climate sensitivity parameter of $\sim$1 K/Wm$^{-2}$ \citep{IPCC_Chap9_2013}), falling into a full glaciation. However, a glaciated early Earth is in contradiction with the evidence for liquid water and the temperate/warm climates discussed in the previous section. Moreover, once the Earth is fully ice-covered with a high surface albedo, increasing the solar constant to its present value would not be enough to exit from the snowball state \citep{sellers69,budyko69,hoffman17}. The faint Young Sun problem becomes a paradox if we assume that the early Earth's atmosphere and continents were the same as today. But there is no argument for such an assumption. Most of the proposed solutions to the faint young Sun problem are based on changes to the atmospheric composition, clouds or land distribution, leading to a stronger greenhouse effect or a lower planetary albedo.

\subsection{CO$_2$}
The first solution to the faint young Sun from \cite{sagan72} was based on a strong greenhouse effect by an ammonia-rich and highly reduced atmosphere. Such an atmosphere is now discarded in favor of a moderately-oxidized, N$_2$- and CO$_2$-rich atmosphere, where the latter would be the dominant greenhouse gas \citep{catling20}. Solving the faint young Sun problem with a high concentration of CO$_2$ is attractive since CO$_2$ is one of the major volatiles released from magmatic degassing \citep{gaillard14}. Its concentration is regulated by the carbonate-silicate cycle \citep{walker81}, which acts as a long-term thermostat on the climate. We discuss the ability of the carbonate-silicate cycle to moderate CO$_2$ at levels required to solve the faint young Sun problem in section 5.

\begin{figure}[h!]
	\includegraphics[width=9.cm]{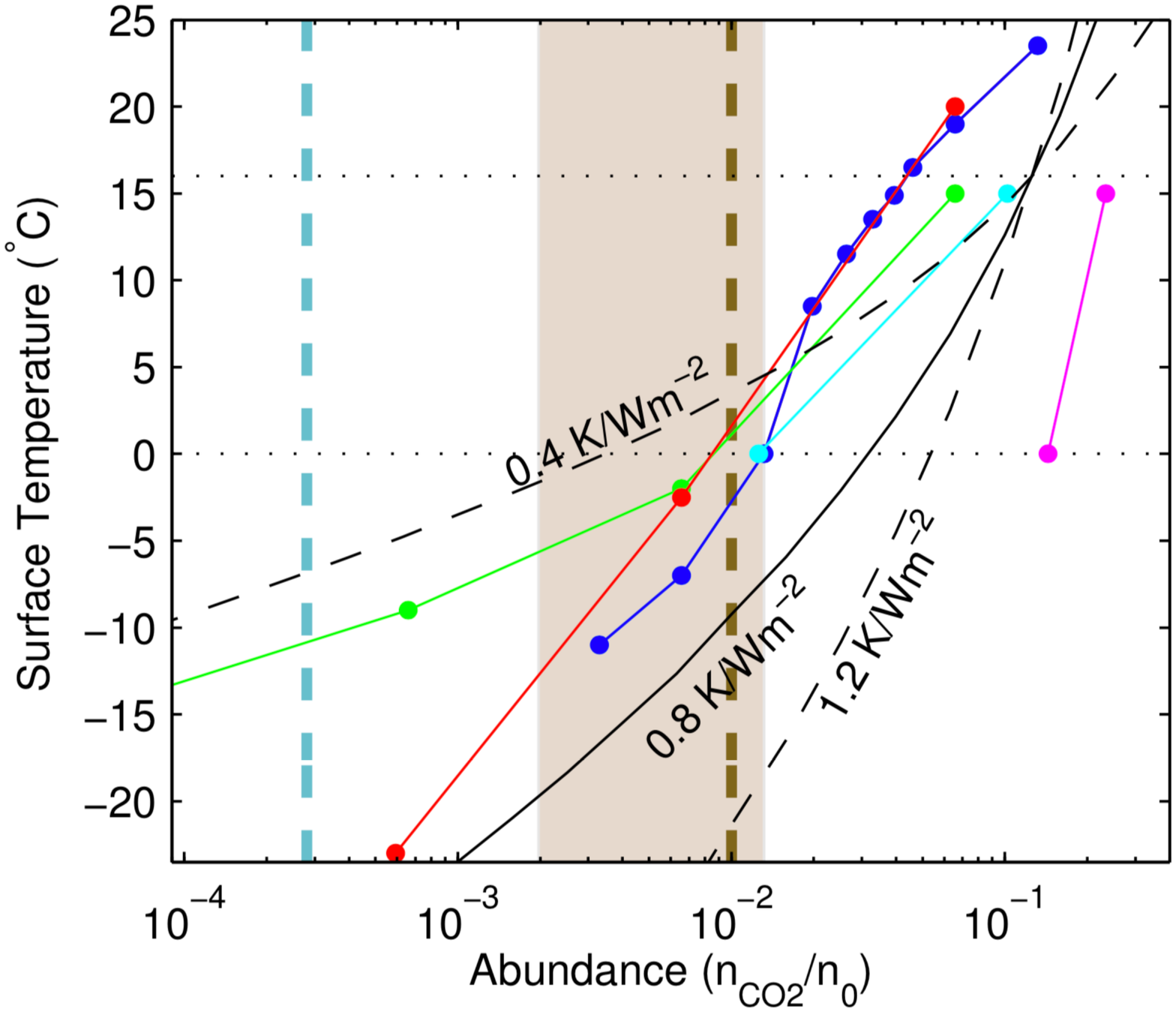}
\caption{Mean surface temperature as a function of CO$_2$ abundance for 0.8S$_0$. The results of the 1D models from \cite{haqq-misra08} (green) and \cite{vonparis08} (cyan) are shown, as well as the 3D models from  \cite{wolf13} (blue),  \cite{charnay13} (red), as well as a 3D oceanic model with a simplified atmospheric model from \cite{kienert12} (magenta). The shaded region shows the CO$_2$ constraint range from \cite{driese11}. The vertical dashed blue and brown lines give the pre-industrial and early Earth guess ($10^{-2}$) abundances of CO$_2$. Figure from \cite{byrne14}.}
\label{figure_1}
\end{figure}

\begin{figure}[h!]
\begin{center} 
	\includegraphics[width=8.5cm]{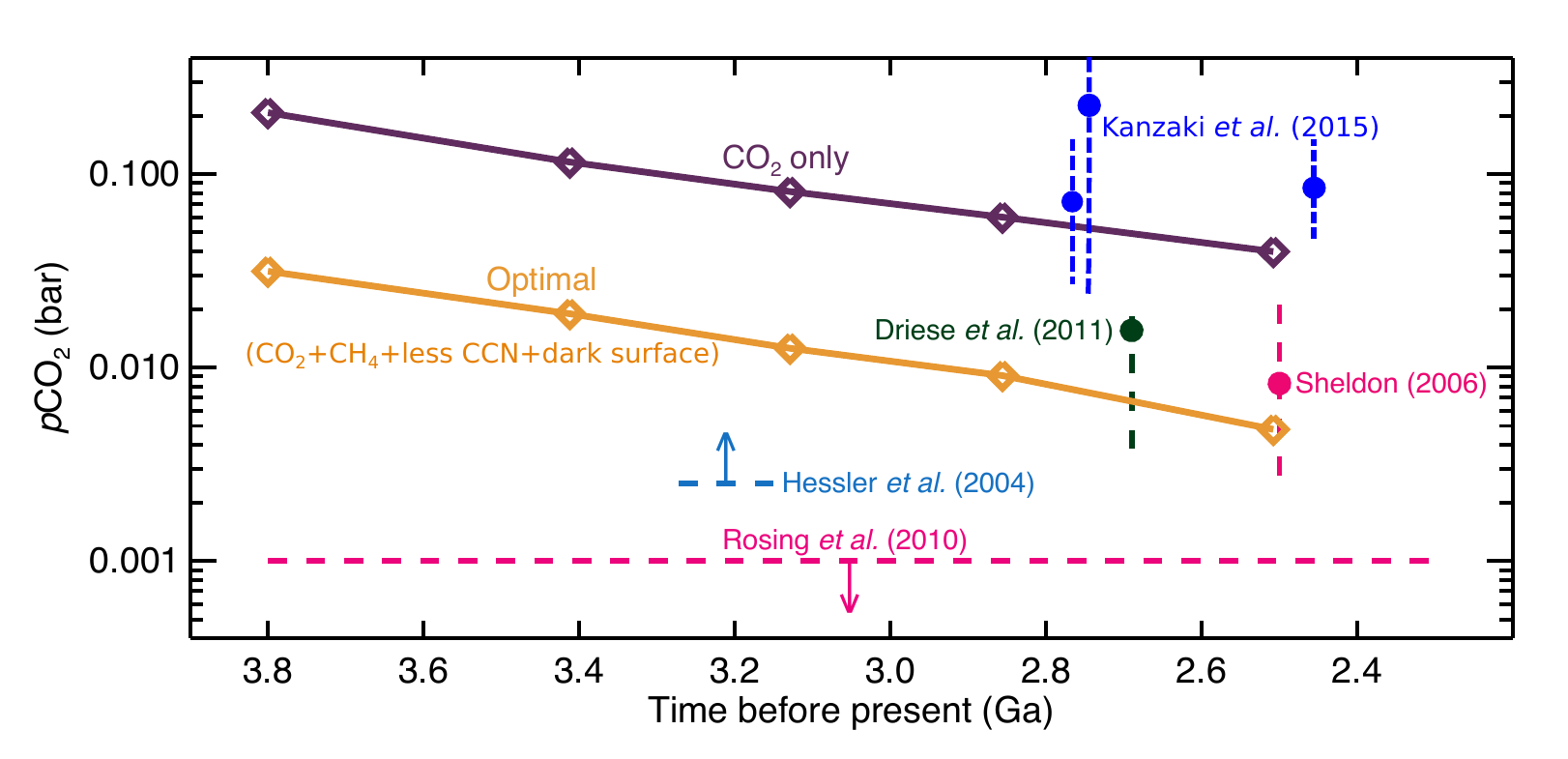}
\end{center} 
\caption{The amount of CO$_2$ needed to maintain global mean surface temperatures of 288 K over the course of the Archean for several different scenarios.  Optimal warming scenarios include 0.1 mbar of CH$_4$, an 18 hour rotation rate, a dark soil surface, and reduced cloud droplet number concentrations.  Figure adapted from \cite{wolf14}.} 
\label{figure_2}
\end{figure}

Previous 1D models showed that $\sim$300 mbar of CO$_2$ is required to maintain a global mean surface temperature of 15$^\circ$C (288 K) during the early Archean and $\sim$100 mbar of CO$_2$ at the late Archean \citep{owen79,kasting84,kiehl87,vonparis08}. 
Such CO$_2$ concentrations exceed the upper limits of $\sim$30 mbar derived from estimates of weathering of paleosols at the end of the Archean \citep{sheldon06,driese11}. 
From these geological constraints and 1D climate modelling results, it has been suggested that other warming processes are needed to keep the early Earth warm. 
\\
\textit{\underline{Lessons from recent studies:}}\\
3D atmospheric GCMs coupled to simple ocean/sea-ice models found that lower amounts of CO$_2$ are needed to maintain global mean surface temperatures of 15$^\circ$C during the Archean (see Fig. \ref{figure_1} and \cite{wolf13,charnay13}). According to the model from \cite{wolf14}, 200 mbar of CO$_2$ is required at 3.8 Ga and 40 mbar at 2.5 Ga (see Fig.  \ref{figure_2}). The case of a warm early Earth with a mean surface temperature of 60-80$^{\circ}$C is achievable with high levels of CO$_2$ around 0.5-1 bar according to the 3D study from \cite{charnay17}, which is significantly less than previous 1D estimations of $\sim$3 bars of CO$_2$ at 3.3 Ga by \cite{kasting06a}. These discrepancies are mostly related to cloud feedbacks (see section 4), leading to warmer climates with 3D models.
The amounts of CO$_2$ required for temperate climates from 3D models are above the geological limits from \cite{sheldon06} and \cite{driese11}.
However, we point out the fact that revised paleosol pCO$_2$ estimates exist, as discussed in the previous section. If the recent pCO$_2$ limits from \cite{kanzaki15} are correct, temperate climates would be achievable 
with just enhanced CO$_2$ (see Fig. \ref{figure_2}). The pCO$_2$ required for an early Earth at 60-80$^{\circ}$C would still remain too high, compatible with only one data point from \cite{kanzaki15} and with the constraint from \cite{lehmer20}.

\subsection{CH$_4$}
Methane has been suggested as an important complement to CO$_2$ to warm the early Earth \citep{kiehl87, pavlov00}. Photochemical models predict that CH$_4$ had a lifetime typically 1000 times longer in the anoxic Archean atmosphere than in the present-day atmosphere \citep{zahnle86,kasting06a}.
Abiotic sources of CH$_4$ like serpentinization in hydrothermal vents could have maintained concentrations up to $\sim$2.5 ppm \citep{tian11,guzman-marmolejo13}. Biogenic CH$_4$ flux from methanogens could have maintained much higher concentrations between 0.1 mbar and 35 mbar \citep{kharecha05, ozaki18, krissansen-totton18b, schwieterman19, sauterey20}. CH$_4$ would have been produced by H$_2$-based methanogens or by fermentors and acetotrophs decomposing the biomass produced by phototrophs (H$_2$-based phototrophs and ferrophototrophs). 
In addition, the fractionation of atmospheric Xenon could be explained by hydrodynamic hydrogen escape during the Hadean/Archean if the H$_2$ mixing ratio was higher than 1\% or if the CH$_4$ mixing ratio was higher than 0.5\% (i.e. 5 mbar for a 1-bar surface pressure) \citep{zahnle19}. Such high concentrations of CH$_4$ produce a strong greenhouse by absorbing thermal radiation at 7–8 $\mu$m, the edge of an atmospheric window for a CO$_2$-H$_2$O atmosphere \citep{kiehl87,pavlov00,haqq-misra08}. 

A limitation appears for high methane concentrations due to the formation of organic hazes. Organic hazes are expected to form when the CH$_4$/CO$_2$ ratio becomes higher than $\sim$0.2 according to photochemical models and experimental data \citep{zerkle12,trainer06}. There is possible isotopic evidence of organic haze formation at 2.7 Ga \citep{zerkle12}.
Hazes are expected to cool the surface by absorbing UV and visible solar radiation, which produces an anti-greenhouse effect \citep{pavlov01, haqq-misra08}. Although, 1D simulations that incorporate the fractal aggregate nature of haze particles suggest that this cooling effect may not be as large as previously thought \citep{wolf10,arney16}. In addition, fractal haze particles act as a UV shield, protecting both life and photolytically unstable reduced gases \citep{wolf10}.

\begin{figure}[h!]
\begin{center} 
	\includegraphics[width=9.cm]{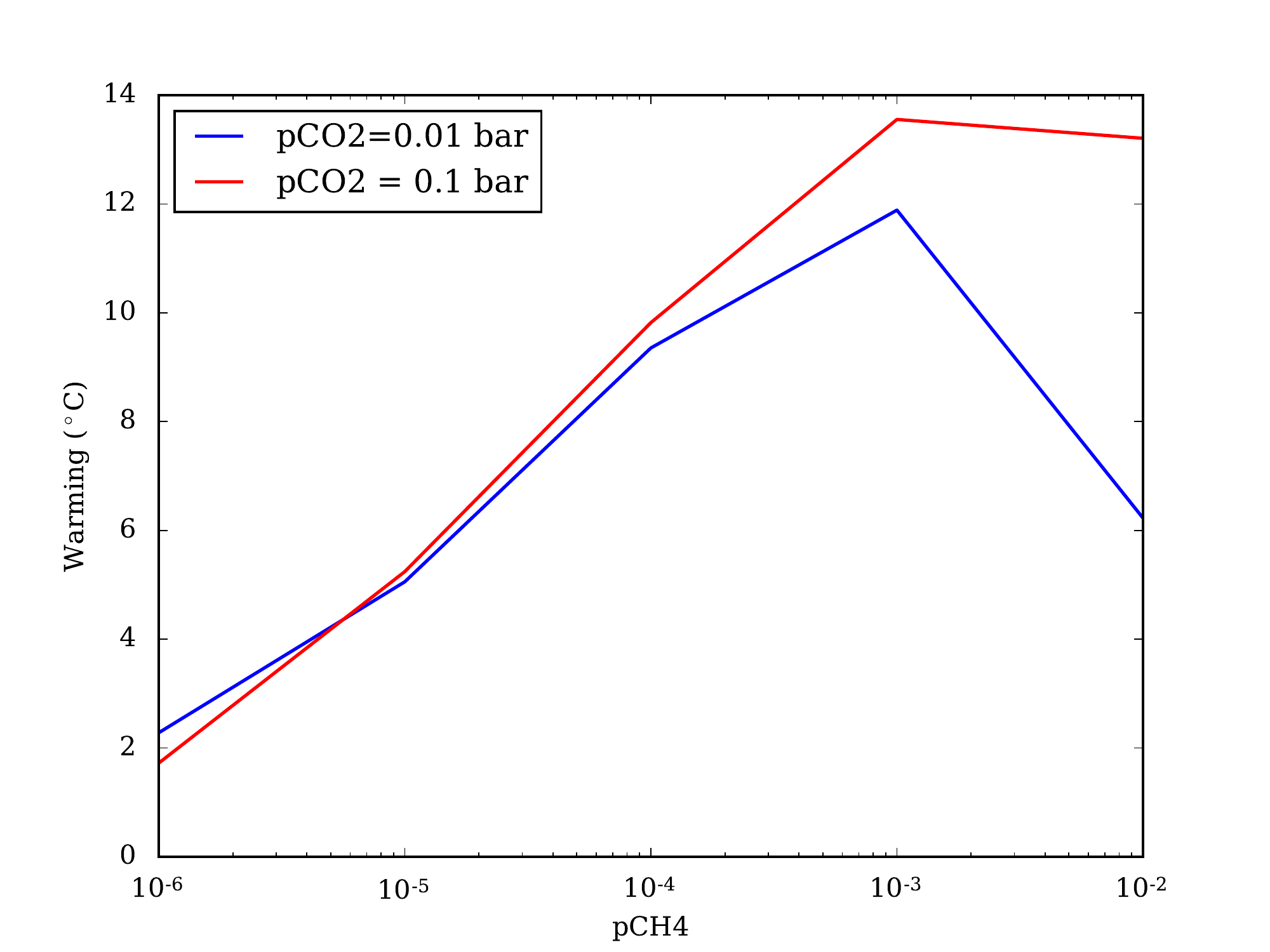}
\end{center} 
\caption{Global mean warming by methane as a function of pCH$_4$ (in bar) for an atmosphere with 0.01 bar (blue) and 0.1 bar (red) of CO$_2$ at 3.8 Ga. Simulations done with the Generic LMD GCM using HITRAN 2008 for CH$_4$ opacity \citep{charnay17}.} 
\label{figure_3}
\end{figure}

Finally, it has been argued that the Great Oxidation Event caused a large decline in atmospheric methane concentration, triggering the Huronian glaciations \citep{pavlov00,kasting06b,goldblatt06}. The apparent synchronicity of the two events at $\sim$2.4 Ga is an additional argument in favour of a methane-rich Archean atmosphere.
\\
\textit{\underline{Lessons from recent studies:}}\\
Fig. \ref{figure_3}  shows the global mean warming by CH$_4$ from a 3D GCM. It reaches up to $\sim$+14 K for pCH$_4$=1 mbar. However, the warming decreases for pCH$_4$$>$1 mbar. That is due to the absorption of near-IR solar radiation by CH$_4$ in the stratosphere, producing an anti-greenhouse effect \citep{byrne14}. That cooling appears in radiative transfer code using HITRAN 2008 or more recent versions. Most of previous 1D studies of the early Earth as \cite{kiehl87,pavlov00,haqq-misra08} used older databases and overestimated the greenhouse effect for high concentrations of CH$_4$. That effect poses a strong limitation to solutions to the faint young Sun problem based primarily on a CH$_4$ greenhouse. 
Despite the limitations at high conceterations, CH$_4$ greenhouse is an excellent complement to CO$_2$ to solve the faint young Sun problem, filling up to 20$\%$ of the radiative forcing deficit \citep{byrne14}. However, it cannot be significant in prebiotic time, meaning that CO$_2$ greenhouse and other abiotic processes likely maintained liquid surface water before the emergence of methanogenesis.

\subsection{Other greenhouse gases (NH$_3$, H$_2$) and atmospheric pressure}
As mentioned previously, \cite{sagan72} proposed an ammonia (NH$_3$) greenhouse as a solution to the faint young Sun problem. However, NH$_3$ is quickly photolyzed and cannot reach concentrations high enough to compensate for the faint young Sun. Yet, UV shielding by photochemical organic hazes could significantly extend the lifetime of NH$_3$ \citep{wolf10}. 3D simulations coupling chemistry and haze microphysics will be required to verify if hazes can adequately protect NH$_3$ from photolysis. In any case, an NH$_3$ greenhouse appears as a limited solution to the faint young Sun problem, requiring a very reduced early atmosphere or high levels of biogenic CH$_4$.

N$_2$ has no electric dipole and interacts weakly with electromagnetic field, making it  a poor greenhouse gas. However, it can enhance the greenhouse effect of others gases by pressure broadening. A high pressure also increases the surface temperature by increasing the moist adiabatic lapse rate, resulting in decreased convective heat transport. \cite{goldblatt09} found that 2-3 times more atmospheric N$_2$ would have produced a warming of +3-8 K, allowing more clement conditions for the early-Earth with a CO$_2$-poor atmosphere. 
At high pressure, N$_2$ and H$_2$ can absorb thermal radiation through collision-induced absorption (CIA). CIA of N$_2$-N$_2$, H$_2$-N$_2$ and N$_2$-CH$_4$ are responsible for most of the greenhouse effect on Titan \citep{mckay91}, while H$_2$-H$_2$ and H$_2$-He play a significant role in the thermal structure of giant planets. \cite{wordsworth13} suggested that the early Earth could have been warmed by N$_2$-H$_2$ CIA. This warming process strongly depends on the amount of N$_2$ and H$_2$. It becomes efficient ($>$+10 K) for a N$_2$ mass column of 2-3 times higher than today and H$_2$ mixing ratio of $\sim$0.1. A model by \cite{tian05}, suggested that the H$_2$ mixing ratio could have been as high as 0.3 in the anoxic early Earth's atmosphere, due to a weaker Jean atmospheric escape. 
\\
\textit{\underline{Lessons from recent studies:}}\\
A recent modelling work by \cite{kuramoto13} shows that atmospheric escape almost reached the diffusion limited regime during this time period. This implies a H$_2$ mixing ratio lower than 0.01 for realistic H$_2$ volcanic fluxes. The existence of detrital magnetite grains in Archean sandstones also suggests a low H$_2$ mixing ratio with pH$_2$ $<$ 10 mbar \citep{kadoya19}.
Moreover, the previous two warming processes, pressure broadening by increased N$_2$ and N$_2$-H$_2$ CIA, both require a N$_2$ mass column 2-3 times higher in the past. Such a high surface pressure is not consistent with geological and isotopic constraints discussed in section 2.5. We conclude that a warming by NH$_3$, N$_2$-H$_2$ or a higher atmospheric pressure probably did not operate during the Archean. The situation might have been different during the Hadean. In particular, the partial pressure of N$_2$ was likely higher in the Hadean when nitrogen fixation was limited to abiotic processes \citep{stueken15,stueken16}. 
An episodic reduced H$_2$-rich atmosphere could also have been formed for 10-20 Myrs after giant impacts caused during the late veneer, by reduction of water by a molten iron impactor \citep{benner19}.

\subsection{A lower albedo due to less continent and cloud feedbacks}

A lower planetary albedo due to a change in the cloud cover and emerged land has been suggested several times as a potential contributor to solve the faint young Sun problem \citep{jenkins93,rosing10}. To fully compensate for a 20-25$\%$ lower insolation, the planetary albedo would have to be $\sim$0.05-0.1 (similar to Mercury or the Moon) compared to 0.29 for the present-day Earth \citep{wild13}. Even by removing of all emerged land and all clouds, such a low value would not be reached because of atmospheric Rayleigh scattering. Still, a change in the cloud cover or thickness and fraction of emerged land could have been part of the solution to the faint young Sun problem.

As explained before, the fraction of emerged land is expected to have been lower during the Archean, likely between 2 and 12 $\%$ of Earth's surface \citep{flament08}. Furthermore, what emergent continents did exist were likely barren of vegetation, and probably were composed of dark basalts, before gradually changing into the lighter-colored soils we see today as they aged.  A reduction of surface albedo would give a maximal radiative forcing of $\sim$+5 W/m$^2$ \citep{goldblatt11b}.

Clouds have a strong impact on the terrestrial radiative budget by reflecting solar radiation and by absorbing thermal radiation. The albedo effect (cooling effect) dominates for lower clouds, such as stratus clouds while the greenhouse effect (warming effect) dominates for upper clouds, such as cirrus clouds. 
\cite{rondanelli10} suggested that the amount of tropical cirrus formed by detrainment from convective clouds would increase under a fainter Sun, enhancing their greenhouse effect and maintaining a clement climate. Such a negative feedback, called the "Iris hypothesis"  \citep{lindzen01}, remains debated \citep{goldblatt11,mauritsen15} and could also be compensated or dominated by other positive cloud feedbacks \citep{bony05,bony15}.
A more plausible hypothesis is that lower clouds were optically thinner during the Archean, owing to the lack of cloud condensation nuclei (CCN) from biological sources. Such a decrease of CCN yields larger cloud particles which are less effective at scattering and precipitate faster, resulting in a decrease of the planetary albedo \citep{rosing10}. The removal of all lower clouds compared to the present-day cover would lead to a radiative forcing of $\sim$25 W/m$^2$  \citep{goldblatt11}.
\\
\textit{\underline{Lessons from recent studies:}}\\
3D GCMs have shown that plausible reductions to the surface albedo and emerged land fraction could have increased the global mean surface temperature by up to +4 K \citep{wolf13,charnay13}.  
They also suggest that a reduction in CCN results in an increase of the cloud radiative forcing of up to $\sim$+12 W/m$^2$, and an increases in the global mean surface temperature of up to 10 K \citep{charnay13, wolf14}. A similar temperate change by CCN was found for the Neoproterozoic by \cite{feulner15}. Independent of the CCN effect, GCMs indicate that a reduced cover of lower clouds and a increased cover of higher clouds for enhanced CO$_2$ and weaker insolation, providing an additional radiative forcing of $\sim$+5 W/m$^2$ (see section 4.2 and \cite{wolf13,charnay13}). 
In conclusion, a change in the land and cloud cover together cannot compensate more than 60$\%$ of the deficit of radiative forcing due to the weaker Archean Sun. A higher amount of greenhouse gases (i.e. CO$_2$ and CH$_4$) is still required. However, a reduction to the cloud and surface albedos may have meaningfully contributed to warming the early Earth, and thus less CO$_2$ and CH$_4$ may have been otherwise required.

\section{New insights from 3D climate models}

\subsection{Testing the proposed warming solutions with 3D GCMs}

While 3D GCMs are notoriously computationally expensive to run, recent gains in computing availability and parallel processing now allow GCMs to be run with sufficient speed to facilitate a wide variety of planetary modelling studies. As such, atmospheric GCMs with simplified ocean/sea-ice components have now been used to thoroughly explore the climate history of the ancient Earth, from the early Archean up through recent snowball Earth epochs \citep{jenkins99, pierrehumbert04, charnay13, wolf13, wolf14,  lehir14, kunze14, teitler14, charnay17, wolf18}. The use of such 3D climate models has allowed us explore the numerous alternative climatological mechanisms for warming the early Earth despite the faint young Sun and described in section 3.

As explained in the former section, these models have shown that enhanced CO$_2$ likely played the major role and could have compensated the fainter Sun with pCO$_2$$\sim$40 mbar at 2.5 Ga. Such a concentration is above the limits from \cite{sheldon06} and \cite{driese11} but compatible with the constraints from \cite{kanzaki15} (see Fig. \ref{figure_2}).Note that the case of a warm early Earth with a mean surface temperature of 60-80$^{\circ}$C is achievable with high levels of CO$_2$ around 0.5-1 bar \citep{charnay17}. However, the pCO$_2$ required in that case is only compatible with one data point from \cite{kanzaki15} (i.e. pCO$_2$=22-700 mbar at 2.75 Ga). CH$_4$ could have been a significant contributor to warm the Archean Earth, increasing the global mean surface temperature by up to +14 K for pCH$_4$=1 mbar (see Fig. \ref{figure_3}  and \citep{wolf13}).

Plausible reductions to the surface albedo or emerged land fraction could have warm the planet by up to +4 K \citep{charnay13,wolf14}. A reduction of emerged land fraction does not necessarily imply a lower planetary albedo because cloud cover tends to decrease above continents compared with over oceans. But this cooling effect is compensated by a stronger evaporation leading to an increase in the amount of water vapour in the atmosphere and thus a stronger greenhouse effect \citep{charnay13}.

GCMs suggest that increases in the global mean surface temperature of up to +10 K are possible to due to reductions in CCN and the resulting feedbacks on clouds. The models predict that there are globally less low clouds (i.e., lower than 5 km), which are in addition optically thinner, for low and middle latitudes with less CCN. However, there are also more low clouds at high latitudes and more high clouds (i.e., higher than 5 km), because of the warmer climate with a more extended Hadley cell. According to \cite{wolf13} and \cite{charnay13}, the warming is dominated by the decrease of low clouds and the shortwave forcing (albedo effect). In the model by \cite{lehir14}, the cloud feedback is dominated by the increase of high clouds and the longwave radiative forcing (greenhouse effect). We note that \cite{lehir14} used a parametrisation for the precipitation rate more sensitive to cloud particle size than \cite{wolf13} and \cite{charnay13}, which may explain the difference.

If the background N$_2$ partial pressure of the Archean was 2 to 3 times greater than today, as postulated by \cite{goldblatt09}, then the Archean may have been warmed by +5 to +10 K for a given amount of CO$_2$ according to these 3D models.  However and as explained in section 2.5, some geological/geochemical analyses indicate that the total nitrogen content of the Archean atmosphere should have been similar or smaller than the present day \citep{som12,marty13,som16}.

The early Earth had a faster rotation rate than the present-day Earth, with a diurnal period possibly as short as 10 hours \citep{zahnle87,williams00,bartlett16}. Changing the rotation would subtly affect horizontal heat transports. A faster rotation rate limits the size of eddies as well as the latitudinal extension and the strength of the Hadley cells. All these changes reduce the efficiency of meridional transport. Under such conditions, the equator-pole thermal gradient is enhanced with a warmer equator and cooler poles, sea ice is more extended and zonal jets are closer to the equator. However this has been shown not to have a significant effect on the overall climate state, with just a small warming of $\sim$+1 K \citep{charnay13,wolf14,lehir14}. Note that these studies use simple ocean models, with no sea-ice transport and with a fixed oceanic heat diffusivity. \cite{charnay13} use a 2-layer ocean model with Ekman transport coupled to surface winds. This allows to more properly simulate meridional heat transport at low latitude, but the model still misses the sea-ice transport. In contrast, simulations by \cite{kienert12} with a 3D oceanic model and a simplified 2D atmospheric model suggest a strong cooling for a fast rotation rate. This is caused by a reduced atmospheric and oceanic heat transport as well as by a change in cloud cover and lapse rate. The latter is likely an artefact of the parametrisation used in the 2D atmospheric model and the cooling is likely overestimated (see the discussion in \cite{lehir14}). However, the change in ocean dynamics may have a non-negligeable impact on the global mean surface temperature. This has to be explored with a full 3D atmosphere-ocean GCM.

Table 1 summarises the different proposed solutions to the faint young Sun problem, the radiative forcing (computed from 1D or 3D models) and the constraints (geological, geochemical or theoretical). Solutions which are compatible with geological/geochemical constraints are in green, possible solutions for which there is only theoretical constraints are in yellow and solutions which are not compatible with the constraints are in red.
While many researchers have sought to find sweeping solutions to the faint young Sun problem, perhaps the paradox is instead solved by a collection of the processes outlined above. Coupled with enhanced CO$_2$, increased CH$_4$, reductions to the cloud albedo, and a dark surface, all can contribute +4 to +14 K of global mean warming. Note that the cumulated radiative forcing (i.e. global warming) is not necessary equal to the sum of each process taken separately. For instance, gases and high ice clouds can have overlapping greenhouse effects.
3D studies now suggest that there are plausible solutions for maintaining temperate climates within geological constraints on surface temperature and CO$_2$, combining these different warming processes.  In particular, \cite{charnay13} and \cite{wolf13} using structurally similar but independently constructed 3D climate models with different boundary conditions (i.e. land distribution), both found that mean surface temperatures similar to that of the present day Earth can be maintained at the end of the Archean (2.5 Ga), provided that the atmosphere had:  1) 40 mbar CO$_2$, 2) 10-20 mbar CO$_2$ and 1-2 mbar CH$_4$ or 3) 5 mbar CO$_2$, 0.1 mbar of CH$_4$, reduced CCN and dark soil surface. The yellow curve in the figure 2 shows the amount of CO$_2$ needed to maintain global mean surface temperatures of 15$^\circ$C considering the case where multiple warming mechanisms are factored in (0.1 mbar of CH$_4$, reduced CCN and dark soil surface). The combination of all these processes is totally plausible for the Archean. For this optimal case, a temperate Earth can be maintained at all times of the Archean with CO$_2$ amounts less than 30 mbar, compatible with constraints from \cite{sheldon06} and \cite{driese11}.

Finally, the similar results obtained from these different 3D GCMs with simplified ocean/sea-ice models, in particular the similar pCO$_2$ required for temperate climates, suggest that the faint young Sun problem can be solved more easily than initially thought. Even the case of a warm early Earth at 60$^\circ$C is more accessible from a purely climate modelling point of view, although not compatible with most of geological constraints on pCO$_2$ and surface pressure.

\begin{table*}
\begin{tabular}{| l | l | l | l |}
\hline
Solutions to the							& Maximal radiative forcing										& Constraints								& References 	for constraints		\\
FYS problem 							& 															& (paleosols or theoretical)  					&  							\\ 
  
\hline

\cellcolor{green}Elevated CO$_2$			&	 \cellcolor{green}+26 Wm$^{-2}$ (for pCO$_2$=10 mbar) 				& pCO$_2$=3–15 mbar  (2.69 Ga)				& \citep{driese11}  				\\   
\cellcolor{green}						&	 \cellcolor{green}+44 Wm$^{-2}$ (for pCO$_2$=60 mbar) 				& pCO$_2$=3–25 mbar  (2.5 Ga)				& \citep{sheldon06}    			\\   

\cline{3-4}

\cellcolor{green}        					&	 \cellcolor{green}											& pCO$_2$=24–140 mbar  (2.77 Ga)			& \citep{kanzaki15}  				\\     					
\cellcolor{green}	 					& 	 \cellcolor{green}\citep{wolf13}									& pCO$_2$=22–700  mbar  (2.75 Ga)			& \citep{kanzaki15}				\\    					
\cellcolor{green}        					&	 \cellcolor{green}\citep{lehir14}									& pCO$_2$=45–140 mbar  (2.46 Ga)			& \citep{kanzaki15} 				\\    	
\cellcolor{green}                             			&	 \cellcolor{green}\citep{byrne14}								& CO$_2$$>$70$\%$  	(2.7 Ga)				& \citep{lehmer20} 				\\     
   
\hline

\cellcolor{green}Elevated CH$_4$   			& \cellcolor{green}+9 Wm$^{-2}$	(for pCH$_4$=1 mbar)				&  pCH$_4$$=$0.01-10 mbar                        		& \citep{sauterey20} 				\\
\cellcolor{green} 						&	 \cellcolor{green}\citep{byrne14}								&  CH$_4$$>$0.5$\%$ ($\sim$3.5 Ga)			& \citep{zahnle19}		   		 \\  
\cellcolor{green}            					&	 \cellcolor{green}\citep{lehir14}									&  CH$_4$:CO$_2$$\sim$0.2 ($\sim$2.6 Ga)		& \citep{zerkle12}				 \\  			
  
\hline

\cellcolor{green}Less emerged land 			&\cellcolor{green}+5 Wm$^{-2}$ (with almost no land)					& Fraction=2-12$\%$   (2.5 Ga)					& \citep{flament08} 				\\  
\cellcolor{green}  						&\cellcolor{green}\citep{goldblatt11}									& Crust volume=60-80$\%$ (3 Ga)				&  \citep{hawkesworth19}			\\  

\hline

\cellcolor{green}Faster rotation 				&\cellcolor{green}$\sim$+0 Wm$^{-2}$ (for P=14h)						& Length of day=21.9$\pm$0.4h (620 Ma)			& 	 \citep{williams00}			\\  
\cellcolor{green} 						&\cellcolor{green}\citep{charnay13}									& Length of day$\sim$13h (3.8 Ga)				&  \citep{bartlett16}				\\    
  
\hline

\cellcolor{yellow!80}Cloud feedbacks  and 	&\cellcolor{yellow!80}$\sim$+5 Wm$^{-2}$ (cloud feedbacks$^*$) 			& Larger droplets (r$\sim$17 $\mu$m)			& \citep{rosing10}				\\  
\cellcolor{yellow!80}less CCN 				&\cellcolor{yellow!80}$\sim$+12 Wm$^{-2}$ (for r=17 $\mu$m) 				& 										& 							\\  
\cellcolor{yellow!80} 						&\cellcolor{yellow!80}\citep{charnay13}								& 										& 							\\  
\cellcolor{yellow!80}   					&\cellcolor{yellow!80}\citep{wolf13,wolf14}								& 										& 							\\  

\hline

\cellcolor{red!60}High surface pressure 		&\cellcolor{red!60}+12.2 Wm$^{-2}$ (for 2$\times$PAL N$_2$) 				& P=0.23$\pm$0.23 bar (2.74 Ga)				& \citep{som16} 				\\	 
\cellcolor{red!60}  				  		&\cellcolor{red!60}\citep{goldblatt09}									& P$<$0.53-1.1 bar	(2.7 Ga)					& \citep{som12}					\\ 

\hline
    
\cellcolor{red!60}N$_2$-H$_2$ warming 		&\cellcolor{red!60} +24 Wm$^{-2}$ (for 3$\times$PAL N$_2$, 10$\%$H$_2$)	& pN$_2$$<$1.1 bar	(3.5-3 Ga)					& \citep{marty13} 				\\ 
\cellcolor{red!60}			 			&\cellcolor{red!60}\citep{wordsworth13}								& pN$_2$$<$1 bar	(3.3 Ga)					& \citep{avice18}				\\    
\cellcolor{red!60}   			 			&\cellcolor{red!60} 												& pH$_2$$<$10 mbar 						& \citep{kadoya19}				\\
\cellcolor{red!60} 			 			&\cellcolor{red!60} 	 											& H$_2$$<$1$\%$  							& \citep{kuramoto13}				\\
  
\hline

\cellcolor{red!60}Elevated NH$_3$ 			&\cellcolor{red!60}+33 Wm$^{-2}$ (for pNH$_3$=10$^{-2}$ mbar) 			& pNH$_3$$\sim$10$^{-5}$ mbar				& \citep{kasting82} 				\\	 
\cellcolor{red!60}  				  		&\cellcolor{red!60}\citep{byrne14}									& Higher values if organic hazes				& \citep{wolf10}					\\ 

\hline
\end{tabular}

\caption{Table of solutions to the Faint Young Sun Problem. The first column lists the different possible solutions. The second column gives the maximal radiative forcing and the corresponding references based on 1D or 3D models. We remind that the faint young Sun implies a deficit of 44 Wm$^{-2}$ at 2.5 Ga and 60 Wm$^{-2}$ at 3.8 Ga. For CO$_2$, we give the forcing for two values of pCO$_2$ consistent with the different constraints. $^*$The change in the cloud radiative forcing is computed for the insolation at 3.8 Ga, between the Archean cloud cover and the present-day cloud cover.
The third and fourth columns show the constraints from paleosols or models for the different solutions with the references (see also table 1 in \cite{catling20}). We indicate here the prominent recent constraints.
Green is for solutions and radiative forcings which are compatible with the constraints, yellow for possible solutions for which there are only theoretical constraints, and red for solutions which are not compatible with the constraints.}
\end{table*}

\subsection{Cloud feedbacks}

Fig. \ref{figure_1}  shows global mean surface temperature as a function of pCO$_2$ from two 1D models  \citep{haqq-misra08, vonparis08}  and two GCMs \citep{wolf13, charnay13} for a 20$\%$ weaker Sun (i.e 3 Ga). While all models predict a pCO$_2$ around 10 mbar for a global mean surface temperature of 0$^\circ$C, the 3D GCMs show a higher climate sensitivity ($\sim$1 K/Wm$^{-2}$ in 3D versus $\sim$0.6 K/Wm$^{-2}$ in 1D), reaching a present-day like mean temperature for half the pCO$_2$ from 1D model. This higher value is mostly due to feedbacks from sea-ice and clouds, not present in 1D models. The climate sensitivity obtained in these 3D studies is similar to the climate sensitivity of GCMs used for the IPCC report, giving a value of 1$\pm$0.5 K/Wm$^{-2}$ \citep{IPCC_Chap9_2013}.

\begin{figure}
	\includegraphics[width=8.5cm]{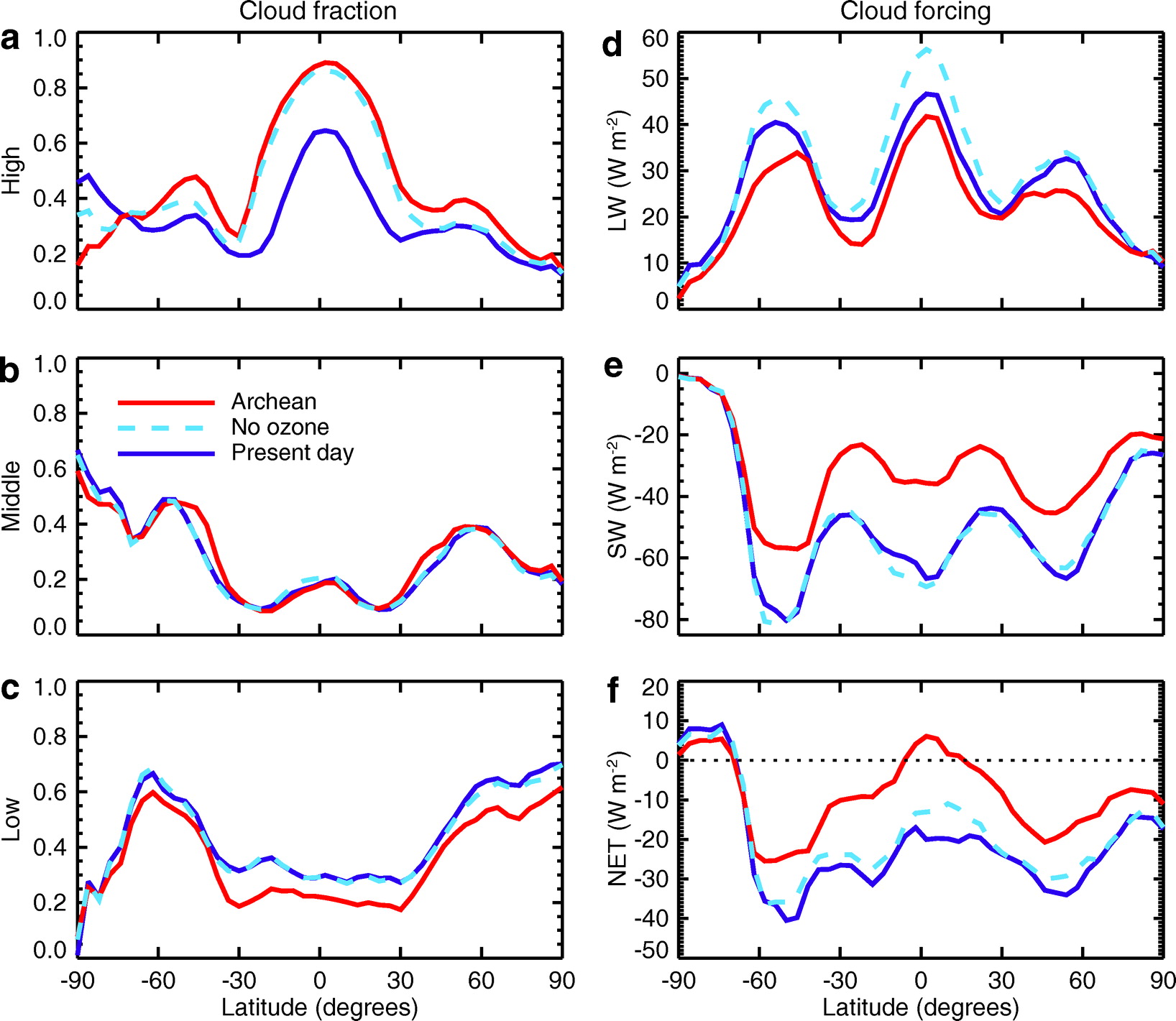}
\caption{Zonal mean cloud properties in 3D atmospheric simulations of the Archean (red), present day (purple), and the present-day atmosphere but with oxygen and ozone removed (dashed light blue). (a) Vertically integrated high cloud fraction. (b) Vertically integrated middle cloud fraction. (c) Vertically integrated low cloud fraction. (d) Longwave cloud forcing. (e) Shortwave cloud forcing. (f) Net cloud forcing. Figure from \cite{wolf13}.} 
\label{figure_4}
\end{figure}

\begin{figure}
	\includegraphics[width=8.5cm]{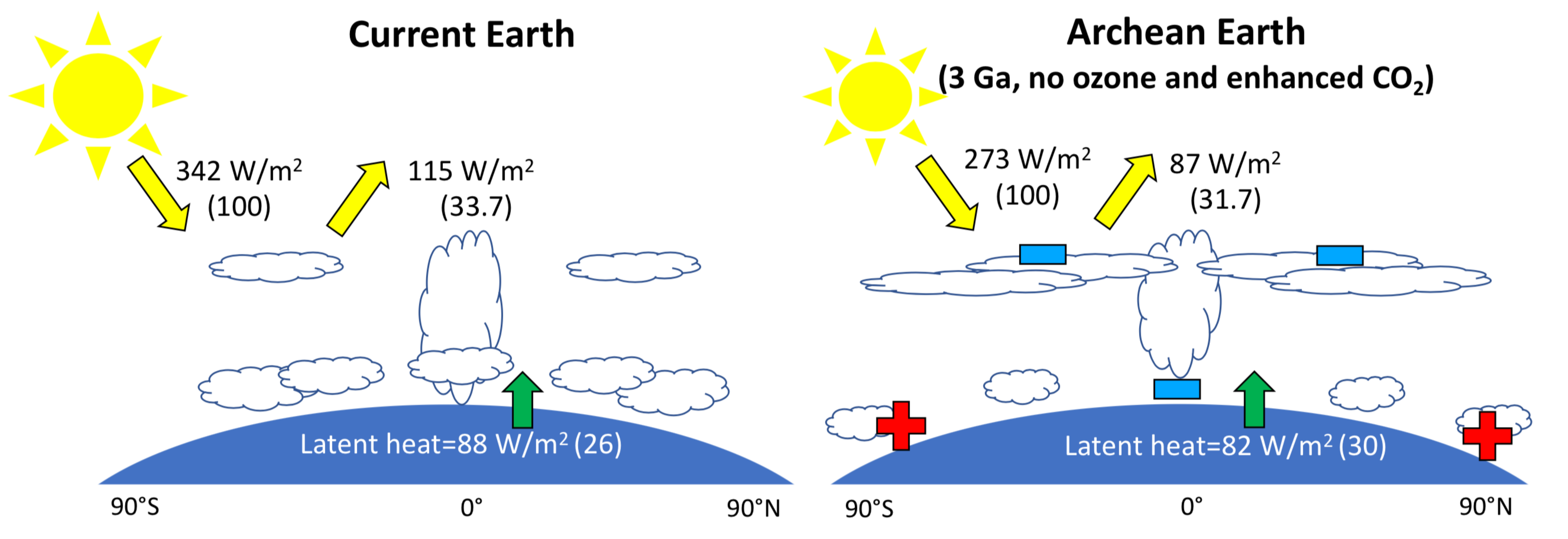}
\caption{Illustration of the change in cloud cover for the Archean Earth. The left panel shows the current Earth with global mean insolation at top of the atmosphere and reflected radiation (yellow arrows), latent heat flux (red arrow) and cloud cover. The fluxes in percent of the global mean insolation are given in brackets. The right panel shows the Archean Earth, with the insolation at 3 Ga, no ozone and enhanced CO$_2$ (i.e. 60mbar) for the same global mean surface temperature as current Earth (287.9 K). Red + indicate the warming of high latitudes and blue - indicate  the cooling of equatorial regions and the upper troposphere compared to the current Earth. The Archean Earth has less lower clouds and more upper clouds. Values are taken from \cite{wolf13}.} 
\label{figure_5}
\end{figure}

For a similar global mean surface temperature, a high pCO$_2$ and a fainter Sun result in a weaker equator-pole temperature gradient, a lower surface evaporation rate and a higher tropopause. The evaporation is reduced by around 7$\%$ for the Archean Earth with present-day like temperatures \citep{wolf13}. Moreover, enhanced CO$_2$ reduces the radiative cooling of low clouds and thus reduces their efficiency to form. Both the weaker evaporation rate and the weaker radiative cooling lead to a reduced fraction of low clouds for the Archean Earth \citep{wolf13, charnay13}.
In addition, 3D GCMs predict an increase of the cover and altitude of high clouds, due to a more extended troposphere and a colder tropopause for a CO$_2$-rich atmosphere with no ozone. The decrease of low clouds and the increase of high clouds increases the net cloud radiative forcing. Fig. \ref{figure_4}  and Fig. \ref{figure_5}  illustrates the change in cloud cover and temperature for the Archean Earth. 3D models predict a net cloud radiative forcing increased by around +15 W/m$^2$ for a temperate Archean Earth with 60 mbar of CO$_2$ \citep{wolf13}. This increase also takes into account the reduction of the insolation in the past and cannot be added directly to other forcings to compensate for the deficit of absorbed solar radiation. 
A more proper calculation can be done using the insolation at 3.8 Ga for the shortwave forcing for both the early Earth and the present-day Earth. In this case, the change in the cloud cover increases the net cloud radiative forcing by around +5 W/m$^2$. 
This means that the cloud feedbacks enhance the CO$_2$ warming by $\sim$11$\%$. If we use instead the current insolation for the shortwave forcing (what will be useful for the following comparison with IPCC simulations), we found an increase of the net cloud radiative forcing by around +7 W/m$^2$, or $\sim$16$\%$ of the CO$_2$ warming.

A similar positive cloud feedback due to high clouds and a reduced cover of low clouds for increasing pCO$_2$ is present in most of 3D GCMs used for climate change (see \cite{IPCC_Chap7_2013} and references therein).
To compare the cloud feedbacks for the Archean Earth (for which the reduced insolation is balanced by enhanced CO$_2$) to GCMs used for current climate change, it is appropriate to consider rapid adjustments, that is after the CO$_2$ is increased but before the ocean temperatures have fully adjusted. This avoids to include contributions due to the change in the sea surface temperature. \cite{zelinka13} found that the net cloud forcing increases by $\sim$1.1 W/m$^2$ in simulations from 5 GCMs after an abrupt quadrupling of CO$_2$. Since the forcing for 4$\times$CO$_2$ is around 7.4 W/m$^2$ \citep{IPCC_Chap7_2013}, cloud feedbacks enhance the CO$_2$ warming by $\sim$15$\%$, similar to our estimate for the Archean Earth using current insolation. 
We conclude that the cloud feedbacks in 3D simulations of the Archean Earth are similar and consistent with 3D simulations for anthropogenic climate change. Note that the spread between different climate models is large, in particular for the cloud forcing which remain one of the largest sources of uncertainty in climate modelling. 

The effects of cloud feedbacks is particularly strong for warm/hot climates. Archean climates with mean surface temperature around 60$^\circ$C are obtained with $\sim$1 bar of CO$_2$ according to the 3D model of \cite{charnay17}. This is 3 times less than predictions from the 1D model of \cite{kasting06a}. This discrepancy mostly comes from the strong decrease of lower clouds for high levels of CO$_2$. 
The net cloud forcing is increased by $\sim$+10-15 W/m$^2$ for a warm climate (pCO$_2$=1 bar) compared to a temperate climate (pCO$_2$=0.1 bar) \citep{charnay17}. This corresponds to an increase of $\sim$+25 W/m$^2$ compared to the present-day cloud forcing. Under high pCO$_2$, increasing the size of cloud particles has a limited effect ($<$+3 W/m$^2$), suggesting that the planet has almost reached the maximal cloud forcing related to the reduction of low clouds.

\subsection{The case of cold climates with waterbelts}

\begin{figure}[h!]
\begin{center} 
	\includegraphics[width=8.5cm]{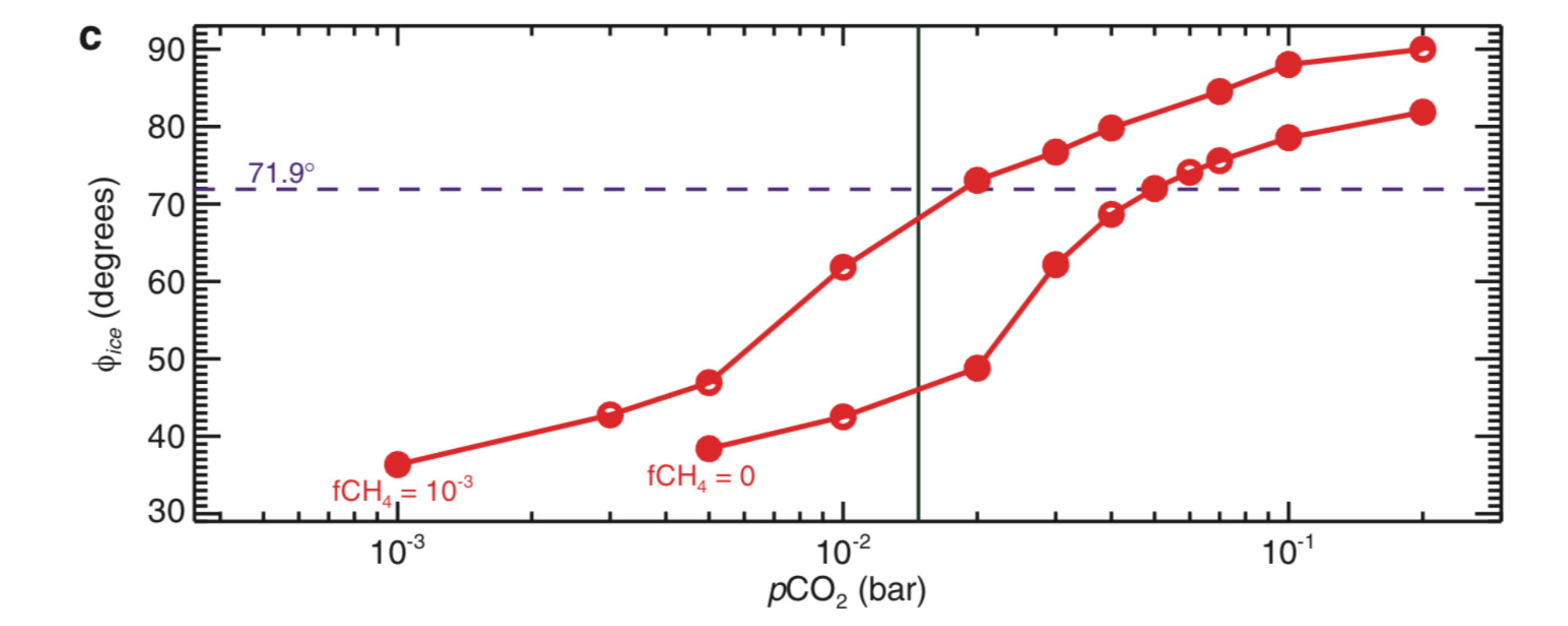}
\end{center} 
\caption{Annual and hemispheric mean sea ice margin ($\phi_{ice}$) as a function of pCO$_2$ from 3D simulations with present-day continents, the Archean Sun at 3.Ga and with or without methane. Purple dashed lines indicate the
sea-ice extent for present-day climate. The green line indicates the pCO$_2$ estimation by \cite{driese11}. Figure from \cite{wolf13}.} 
\label{figure_6}
\end{figure}

With fairly small CO$_2$ amounts ($<$ 1 mbar), 3D models with simplified ocean predict that the early Earth would have been cold (T $<$ 0$^\circ$C) but would have avoided complete glaciation, instead maintaining an equatorial ice-free waterbelt \citep{charnay13, wolf13, lehir14, kunze14, teitler14}. For these cold climates, the mean surface temperature is lower than predicted by 1D models (see for instance Fig. \ref{figure_1}). Fig. \ref{figure_6} shows the mean latitudinal extent of polar sea ice as a function of pCO$_2$ from 3D simulations by \cite{wolf13} at 3 Ga. Sea ice can extend toward the equator down to 37$^\circ$N/S without triggering a runaway glaciation. In the model by \cite{charnay13}, sea ice can extend down to 25$^\circ$N/S. In this model, equatorial water belts are stabilized by a cloud feedback, with a reduction of tropical low clouds due to a subsidence occurring at the ice line. \cite{abbot11} also found a stronger subsidence at the tropics in the case of a waterbelt state. In their model, this induces efficient evaporation of the highly reflective snow (albedo$\sim$0.8) in the tropics, leaving bare sea ice which is less reflective (albedo$\sim$0.4-0.5) and which stabilizes the waterbelt state against runaway glaciation.
According to these studies, at least some surface liquid water, and thus habitable conditions, could have been maintained with only a minimal CO$_2$-CH$_4$ greenhouse. 
If real, such a resistance against full glaciation would mitigate the faint young Sun problem.

However, the stability of cold climates with equatorial waterbelts has been questioned by \cite{voigt12}. In the case of the Neoproterozoic glaciations and using a 3D atmosphere-ocean GCM, they showed  that sea-ice transport plays a major role in the triggering of full glaciation. Glaciation is initiated with 100 times more CO$_2$ when sea-ice transport is included.
We conclude that 3D GCMs with simple ocean/sea-ice models likely overestimate the stability of waterbelts. Archean cold climates with waterbelts should therefore not be considered as robust solutions to the faint young Sun problem unless their stability is demonstrated with full atmosphere-ocean GCMs.

\section{The role of biogeochemical cycles}

\subsection{The carbon cycle as a key for solving the faint young Sun problem}

3D climate modelling with simplified ocean/sea-ice components reveals that temperate climates with present-day like temperatures can be maintained for the late Archean with enhanced CO$_2$, potentially helped by other warming processes.
The carbon cycle and its ability to regulate the climate by the carbonate-silicate cycle thus appears to be a key for solving the faint young Sun problem.

Modelling of the carbon cycle of the early Earth by \cite{sleep01a} and \cite{zahnle02} suggested that the Archean was very cold unless another strong greenhouse gas was present. They also suggested that the Hadean was likely fully ice-covered because of the weathering of impact ejecta, particularly during the Late Heavy Bombardment (LHB)  \citep{gomes05,bottke17}. Impact ejecta are indeed easily weathered when falling in the ocean, consuming CO$_2$. \cite{sleep01a} and \cite{zahnle02} highlighted the importance of the seafloor weathering, caused by the reaction of seawater with the oceanic crust in low-temperature, off-axis, hydrothermal systems \citep{brady97,coogan13,coogan15}. This CO$_2$ sink would have been efficient on the early Earth with little emerged land and a high oceanic crust spreading rate. 
However, calculations of seafloor weathering by \cite{sleep01a} and \cite{zahnle02} did not take into account possible dependence on ocean chemistry, pH and oceanic temperature, and likely overestimated it. Analyses by \cite{brady97,coogan13,coogan15} suggest that the rate of basalt dissolution and pore-space carbonate precipitation depends on bottom water temperature and/or seawater composition. In addition, reverse weathering induced by authigenic clay formation could have favoured high pCO$_2$ on the early Earth \citep{isson18}.

\begin{figure}
	\includegraphics[width=8.5cm]{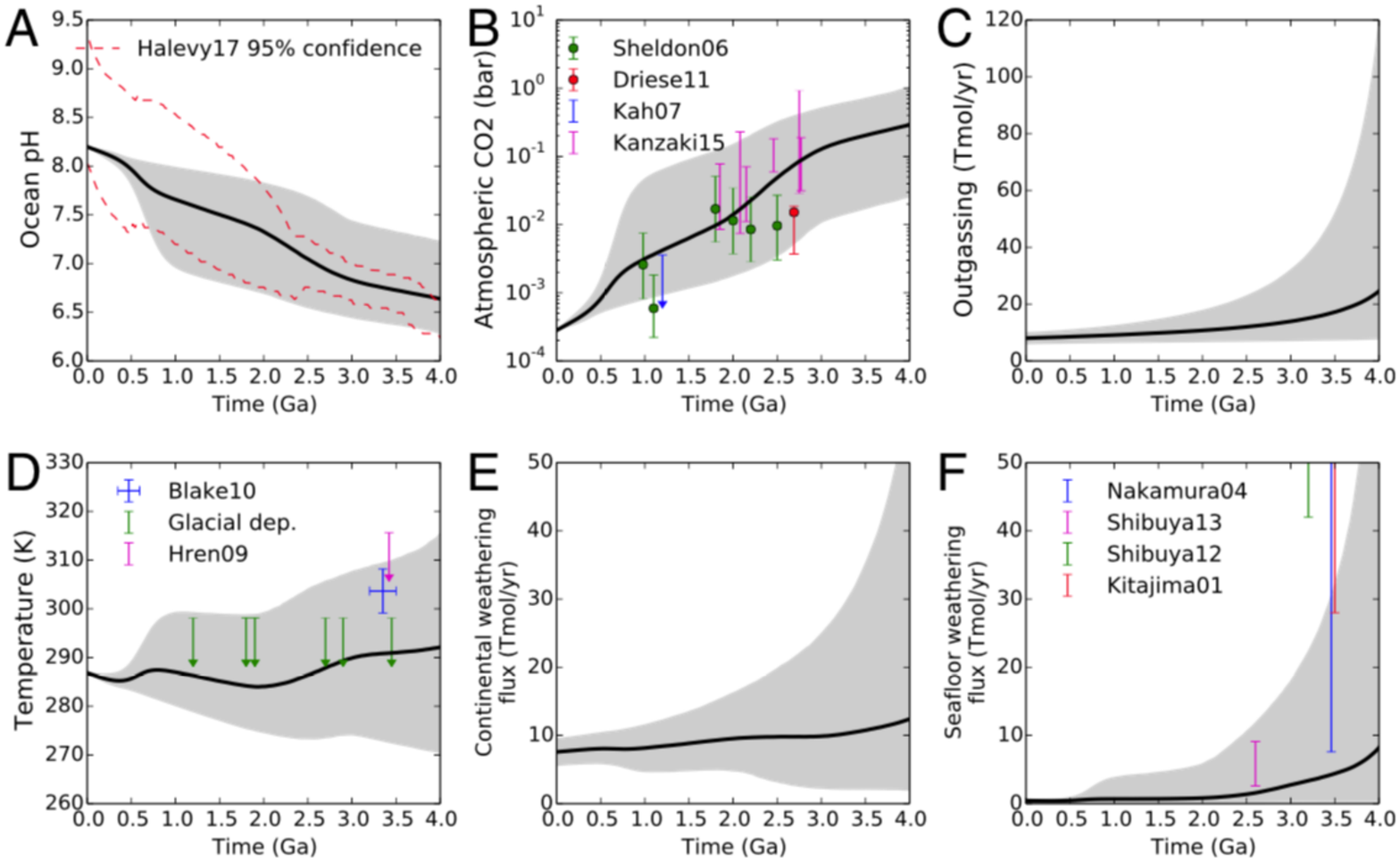}
\caption{Evolution of pH (A), pCO$_2$ (B), global outgassing flux (C), mean surface temperature (D), continental silicate weathering flux (E) and seafloor weathering flux (F) over Earth's history. Gray shaded regions represent 95$\%$ confidence intervals, and black lines are the median outputs of the carbon-cycle model from \cite{krissansen-totton18}. Some geological/geochemical proxies for temperature, pCO$_2$ and seafloor weathering are indicated. Figure from \cite{krissansen-totton18}.} 
\label{figure_7}
\end{figure}

More recent modelling works by \cite{charnay17} and \cite{krissansen-totton18}, using updated climate models and parametrizations of seafloor weathering with composition/temperature dependence, revealed an efficient regulation of the carbon cycle maintaining temperate conditions for the early Earth (see Fig. \ref{figure_7}). 
Seafloor weathering flux in these models is of a similar magnitude or higher than continental weathering flux. According to these studies, warm early oceans at 60$^\circ$C can not be maintained because of the strong negative temperature feedback.
\cite{charnay17} re-evaluated the long-term effect of impacts on the carbon cycle during the Late Heavy Bombardment (whose existence is still debated, see for instance \cite{boehnke16}). They found that the weathering of ejecta would have strongly decreased the partial pressure of CO$_2$ leading to cold climates but not necessary to a snowball Earth during all that period.

In conclusion, these recent modelling studies suggest that the early Earth's climate was likely temperate, except during the Late Heavy Bombardment, and regulated by the carbon-cycle, without necessary requiring additional greenhouse gas or warming process.

\subsection{The biosphere and life feedbacks}
A fascinating topic is the role that life played in the maintenance of Earth's habitability. Some researchers have speculated that the long-term stability of Earth's climate is aided by feedbacks involving life itself \citep{lovelock74,schwartzman89,lenton98,lenton01}. In particular, several possible solutions to the faint young Sun problem rely on the biosphere and life feedbacks, for instance the production of CH$_4$ by methanogens or the reduction of CCN from biological sources. In contrast, metabolic or ecological evolution could destabilize the carbon cycle or the radiative balance, inducing glacial events and mass extinctions. For instance, the Huronian glaciations, the Neoproterozoic glaciations and the Ordovician glaciations could be related to the development of new species (i.e. cyanobacteria, algae, fungi and plants respectively)  \citep{heckman01,lenton12,feulner15}. 
The C isotopic fractionation from ancient rocks suggests the presence of a productive biosphere, likely due to the early emergence of photosynthesis (oxygenic or anoxygenic) \citep{nisbet01,krissansen-Totton15}. This potentially implies a significant impact of the early biosphere on the carbon cycle, but also on the nitrogen cycle (affecting the surface pressure, see \cite{stueken16}) and other biogeochemical cycles (in particular sulfur and phosphorous).
The question about life feedbacks is also fundamental for exoplanets, concerning the habitability and the search for biosignatures \citep{chopra16}. To be tested, it requires global planet models simulating the interaction between ecosystems, biogeochemical cycles and the climate.

A recent modelling work following this approach by \cite{sauterey20} points to the efficient methane production by primitive methanogenic ecosystems. They appear as a robust contributor to solve the faint young Sun problem. However, the methane greenhouse effect enhances rock and seafloor weathering, decreasing pCO$_2$. At equilibrium, this negative feedbacks by the carbon cycle compensates for around 50$\%$ of the methane warming. For some conditions, the decrease of pCO$_2$ by the carbon cycle feedback can lead to the formation of organic hazes (for CH$_4$:CO$_2$$\geq$0.2), cooling the Earth and triggering glaciations \citep{kanzaki18}. 
Finally, the late appearance of methanotrophs, consuming methane once pCO$_2$ was reduced, could also have triggered glacial events \citep{sauterey20}. These are examples of climate destabilization by life with primitive ecosystems. Additional work is needed to investigate all the possible implications of the early biosphere on the climate at different ages.

\section{Conclusions and perspectives}
A few years ago, the faint young Sun problem appeared to be very challenging despite the various proposed solutions: "All of these solutions present considerable difficulties, however, so the faint young Sun problem cannot be regarded as solved." \citep{feulner12}. 1D atmospheric models failed to produce present-day like mean surface temperate when satisfying the geological/geochemical constraints on CO$_2$. In addition, carbon cycle models were not able to maintain temperate climates during the Archean and the Hadean with only CO$_2$ as greenhouse gas.

Major progress concerning the faint young Sun problem has been made during the last decade, in particular with 3D atmospheric GCMs coupled to simplified ocean/sea-ice models which have allowed comprehensive testing of a variety of proposed solutions  \citep{charnay13, wolf13, wolf14, lehir14, kunze14, teitler14, charnay17}. These 3D models overcame the limitations inherent to previous 1D models, taking into account fundamental climate feedbacks (clouds and sea-ice), atmospheric/oceanic heat transport and land distribution, although they do not capture the dynamics of sea-ice and the oceans. They suggest that a temperate early Earth is easier to achieve than previously thought. More precisely, around 200 mbar of CO$_2$ is required at 3.8 Ga to reach present-day mean temperate and 40 mbar at 2.5 Ga, assuming no change in Earth's continents and rotation rate. This is around half of that estimated by 1D models. This discrepancy is mostly due to cloud feedbacks with a reduction of low clouds and an increase of high clouds for enhanced CO$_2$ and reduced insolation. Similar cloud feedbacks are predicted with 3D models used for climate future projections \citep{IPCC_Chap7_2013,schneider19}. Unfortunately, the cloud response which may have helped to warm the early Earth, may worsen the current anthropogenic climate change.

3D studies also revealed that temperate climates can be reached for the late Archean with modest levels of CO$_2$ (down to 5 mbar), if combined with realistic levels of CH$_4$, reduced CCN and less emerged lands. 
CH$_4$ appears as an excellent complement to CO$_2$, although its warming effect saturates at pCH$_4$$\sim$1 mbar (corresponding to a warming of $\sim$+14 K). Ecosystem models coupled to atmospheric models suggest that CH$_4$ concentration of the order of 0.1-1 mbar could have been reached during the Archean \citep{ozaki18,sauterey20}. The reduction of CCN appears as another robust mechanism producing optically thinner low clouds and a significant global warming (up to $\sim$+10 K). The reduction of the fraction of emerged land has a modest but non-negligeable effect  (up to $\sim$+4 K). All these changes (biogenic CH$_4$, reduced CCN and emerged land) are plausible for the Archean Earth. 
In addition, recent geological constraints suggest higher levels of CO$_2$ for the late Archean, generally $\sim$20-140 mbar \citep{kanzaki15} or potentially higher \citep{lehmer20}. Such higher values of CO$_2$ mitigate the faint young Sun problem, since enhanced CO$_2$ alone can then be a sufficient solution. According to recent modelling studies, these levels of CO$_2$ could have been maintained by the carbon cycle, producing an efficient climate regulation \citep{charnay17,krissansen-totton18}. Finally, constraints of the barometric pressure and the N$_2$ abundance suggest that the surface pressure was likely lower than today during the Archean, ruling out solutions involving  a higher surface pressure and N$_2$-H$_2$ CIA.

In conclusion, the 3D atmospheric GCMs combined with carbon-cycle models suggest that the faint young Sun problem can be solved with higher levels of CO$_2$, consistent with the most up to date constraints on pCO$_2$, potentially help by additional warming processes (i.e. biological CH$_4$, less emerged land and less CCN). The case of a temperate early Earth does not appear very problematic anymore. The case of a warm ($\sim$60-80$^\circ$C) early Earth, which is heavily debated, remains challenging. 3D models showed that it requires less greenhouse gases than previously thought, but still exceeding most of the recent constraints on CO$_2$ and on the barometric pressure.

Although we believe that the faint young Sun problem has essentially been solved, new geological and geochemical constraints on the atmospheric composition, pressure and surface temperature are required to get a clear picture of the early Earth's climate. Additional modelling work is also needed, in particular for quantifying the impact of ocean transport and for testing the stability of cold waterbelt states, with atmospheric GCMs fully coupled to 3D oceanic models including sea-ice transport. Such 3D atmosphere-ocean GCMs for the early Earth are a great challenge in term of modelling but would have numerous applications for past climates and for exoplanets. New simulations for low atmospheric surface pressures are also necessary to analyse the changes in terms of heat transport and cloud cover, and to estimate the pCO$_2$ required to avoid glaciation. 
More generally, a major challenge is to understand the co-evolution of life and the environment on Earth. In terms of modelling, this requires to simulate the biogeochemical cycles and the couplings between the different components of the Earth system (ocean, atmosphere, biosphere, land and interior). Recent modelling studies show promising perspectives for the early Earth in this context (see for instance \cite{stueken16,krissansen-totton18,sauterey20}).

Finally, the faint young Sun problem has stimulated a lot of research and new ideas in the fields of solar physics, geology, climate science, planetary science, biology, in particular concerning the interactions between the different components of the Earth system. With its different stellar flux, rotation rate, land distribution, atmosphere and biosphere, we can view the early Earth has another habitable world or even multiple habitable worlds different from our modern Earth. The early Earth appears as a fantastic laboratory to study processes controlling the climate, the atmospheric evolution and the habitability of exoplanets. We can dream that in return, the characterization of terrestrial exoplanets will shed light on these processes and on the environmental conditions which allowed the emergence of life on Earth.

\begin{acknowledgements}
We thank the two anonymous reviewers for comments that improved the manuscript. 
B. Charnay acknowledges financial support from the Programme National de Planétologie (PNP) of CNRS/INSU, co-funded by CNES.
\end{acknowledgements}

%
%



\nocite{*}
\bibliographystyle{agu}

\end{document}